\documentclass[aps,reprint,superscriptaddress,floatfix,pra]{revtex4-2}
\usepackage{graphicx,amsmath,amssymb,amsfonts}
\usepackage{subcaption}
\usepackage{appendix,mathtools}
\usepackage{leftindex}
\usepackage[breaklinks=true,colorlinks,citecolor=blue,linkcolor=blue,urlcolor=blue]{hyperref}
\usepackage[all]{hypcap}
\usepackage{newtxtext,newtxmath}
\usepackage{natbib}
\usepackage{multirow}
\usepackage{booktabs}
\usepackage{bbold}
\usepackage{bm}
\usepackage{anyfontsize}
\usepackage{derivative}
\usepackage[dvipsnames]{xcolor}
\usepackage{ragged2e}
\usepackage{orcidlink}
\usepackage{physics}
\usepackage[percent]{overpic}  

\DeclareFontFamily{U}{BOONDOX-calo}{\skewchar\font=45 }
\DeclareFontShape{U}{BOONDOX-calo}{m}{n}{
  <-> s*[1.05] BOONDOX-r-calo}{}
\DeclareFontShape{U}{BOONDOX-calo}{b}{n}{
  <-> s*[1.05] BOONDOX-b-calo}{}
\DeclareMathAlphabet{\mathcalboondox}{U}{BOONDOX-calo}{m}{n}
\SetMathAlphabet{\mathcalboondox}{bold}{U}{BOONDOX-calo}{b}{n}
\DeclareMathAlphabet{\mathbcalboondox}{U}{BOONDOX-calo}{b}{n}
\begin{document}
\title{Delta-Corrected Shot Reduction for Bypassing Optimization Plateaus in Variational Quantum Eigensolvers}
\author{Jamie Evander Ghozaly\orcidlink{0009-0007-3634-8024}}
\email{jamieevanderghozaly@mail.ugm.ac.id} 
\affiliation{Research Center for Quantum Physics, National Research and Innovation Agency (BRIN), South Tangerang 15314, Indonesia}
\affiliation{Department of Physics, Faculty of Science and Mathematics, Universitas Gadjah Mada, Bulaksumur, Special Region of Yogyakarta, 55281, Indonesia}
\author{Pekik Nurwantoro}
\email{pekik@ugm.ac.id}
\affiliation{Department of Physics, Faculty of Science and Mathematics, Universitas Gadjah Mada, Bulaksumur, Special Region of Yogyakarta, 55281, Indonesia}
\author{Gagus Ketut Sunnardianto\orcidlink{0000-0002-2781-0076}}
\email{gagu001@brin.go.id}
\affiliation{Research Center for Quantum Physics, National Research and Innovation Agency (BRIN), South Tangerang 15314, Indonesia}
\author{Yanoar P. Sarwono\orcidlink{0000-0001-8936-3737}}
\email{yano001@brin.go.id}
\affiliation{Research Center for Quantum Physics, National Research and Innovation Agency (BRIN), South Tangerang 15314, Indonesia}
\author{Leong-Chuan Kwek \orcidlink{0000-0002-0879-0591}}
\email[Corresponding author: ]{kwekleongchuan@nus.edu.sg}
\affiliation{Centre for Quantum Technologies, National University of Singapore, Singapore 117543, Singapore}
\affiliation{National Institute of Education, Nanyang Technological University, 1 Nanyang Walk, Singapore 637616}
\affiliation{MajuLab, CNRS-UNS-NUS-NTU International Joint Research Unit, UMI 3654, Singapore}
\affiliation{Quantum Science and Engineering Centre, Nanyang Technological University, Singapore}

\begin{abstract}
The variational quantum eigensolver (VQE) remains a cornerstone hybrid quantum-classical algorithm for quantum simulation, yet its scalability is fundamentally constrained by heavy measurement overheads. While strategies such as variance-preserved shot reduction (VPSR) mitigates this cost, they suffer from convergence instability and optimization stagnation caused by finite-sample bias in limited pilot shots. To overcome this limitation, we introduce delta-corrected shot reduction (DCSR), an analytically enhanced shot allocation strategy that rigorously mitigates finite-sample instability. Utilizing the multivariate delta method, we derive explicit higher-order bias correction terms for the empirical reduction factor. Both theoretical analysis and numerical simulations reveal that incorporating higher-order statistical moments, such as skewness and kurtosis, provides the gradient resolution necessary to bypass noise-induced barren plateaus. DSCR significantly accelerates convergence compared to standard VPSR across diverse noise and resource regimes while matching or decreasing cumulative measurement costs, establishing a mathematically rigorous framework for variance-stabilized shot allocation on near-term hardware.
\end{abstract}
\date{\today}
\maketitle
\section{Introduction}
\label{sec:int}
Hybrid quantum-classical algorithms have emerged as one of the principal computational paradigms for noisy intermediate-scale quantum (NISQ) devices \cite{preskill_quantum_2018, bharti_noisy_2022, bharti_noisy_2022-1, lau_nisq_2022, peruzzo_variational_2014}. Among them, the variational quantum eigensolver (VQE) has become a leading approach for determining molecular ground-state energies because it combines shallow quantum circuits with classical optimization while exhibiting greater resilience to hardware imperfections than many fully coherent quantum algorithms \cite{peruzzo_variational_2014, tilly_variational_2022, kandala_hardware-efficient_2017, barkoutsos_quantum_2018, romero_strategies_2019, arute_quantum_2019, hakim_quantum_2026, darmawan_improving_2025, iryanti_connecting_2025, mariyanto_fubini-study_2026}. Despite considerable experimental progress, however, the scalability of VQE remains constrained by the large measurement overhead required to estimate expectation values of qubit Hamiltonians \cite{patel_quantum_2025, verteletskyi_measurement_2020, ikhtiarudin_measurement_2026, zhu_optimizing_2024}.

After fermion-to-qubit transformations such as the Jordan-Wigner or Bravyi-Kitaev mappings, an electronic Hamiltonian is expressed as a weighted sum of Pauli operators  \cite{seeley_bravyi-kitaev_2012, mcardle_quantum_2020, cao_quantum_2019, bauer_quantum_2020, kais_introduction_2014, lanyon_towards_2010, kassal_simulating_2011}. Since non-commuting operators must be measured in different bases, the total energy needs to be reconstructed from many independent expectation values obtained through repeated projective measurements. Because each measurement is inherently stochastic, the statistical uncertainty decreases only as the inverse square root of the number of shots, making measurement the dominant computational cost of many VQE calculations. As molecular systems increase in size, this sampling overhead rapidly becomes the primary obstacle to practical quantum simulation. Numerous measurement-allocation strategies have therefore been developed to mitigate this bottleneck. While uniform shot allocation ignores differing statistical contributions, variance-based methods improve efficiency by assigning more shots to groups with larger fluctuations. More recently, variance-preserved shot reduction (VPSR) \cite{zhu_optimizing_2024} further lowered costs by estimating clique variances from a small set of pilot shots and computing a global reduction factor that scales the remaining budget while approximately preserving the target variance. Although VPSR substantially lowers measurement cost, its performance depends critically on the statistical quality of the pilot estimates. In realistic NISQ experiments only a limited number of pilot shots are available, and the empirical variance becomes a noisy estimator of the true variance. Because the VPSR reduction factor is a nonlinear function of these estimated variances, finite-sample fluctuations introduce a systematic bias through Jensen’s inequality \cite{casella_statistical_2024}. Consequently, the measurement budget is frequently under- or over-allocated, leading to inaccurate energy estimates and noisy gradients that slow classical optimization. Under restrictive measurement budgets or elevated hardware noise, this bias often manifests as extended optimization plateaus in which the optimizer loses momentum despite remaining far from the variational minimum. 

In this work, we introduce delta-corrected shot reduction (DCSR), a statistically rigorous framework that analytically removes the finite-sample bias underlying VPSR. Rather than modifying the optimization procedure or introducing heuristic regularization, DCSR directly corrects the measurement-allocation rule through a multivariate delta-method expansion of the empirical reduction factor. Closed-form second-, third-, and fourth-order correction terms are derived, allowing the allocation strategy to account systematically for higher-order statistical moments of the variance estimator while preserving the adaptive structure of VPSR. Extensive numerical simulations of noisy VQE calculations demonstrate that the proposed corrections substantially improve optimization stability across both noise-limited and resource-limited regimes. Higher-order DCSR consistently suppresses the optimization plateaus observed with the original VPSR algorithm, yielding faster convergence while maintaining, or even reducing, the cumulative number of measurements required to reach the target energy. The improvement becomes particularly pronounced under severe measurement constraints, where finite-sample bias is strongest and conventional adaptive shot-allocation methods become unreliable. 

Unlike machine-learning strategies such as Bayesian optimization or online learning, DCSR does not make use of heuristics based on assumptions of prior distributions, nor does it require significant algorithmic training overhead and complex parameter tuning. DCSR is strictly analytical and deterministic in its usage of the Delta Method to mitigate finite-sample bias.

These results identify finite-sample bias as a critical limitation of adaptive measurement allocation and establish DCSR as a general statistical framework for variational quantum algorithms. Because the correction is derived analytically and remains independent of a particular molecular system or ansatz, the method can be seamlessly integrated into broader measurement-reduction protocols across diverse near-term quantum architectures

\section{Theory}
\label{sec:Theory}
The measurement cost of the variational quantum eigensolver (VQE) originates from estimating expectation values of qubit Hamiltonians through repeated projective measurements. After grouping mutually commuting Pauli operators into m measurement cliques, the statistical variance of the estimated energy is
\begin{equation}
    \text{Var}(\bar{E}(\theta)) = \sum_{i=1}^{M} \frac{\sigma_{i}^{2}}{N_{i}},
    \label{equation 1}
\end{equation}
where $\sigma_{i}$ is the is the standard deviation associated with the $i$-th clique and $N_{i}$ denotes the number of allocated measurement shots. Under the constraint $\sum_{i} N_{i} = N$, adaptive shot-allocation schemes seek an optimal distribution of the total measurement budget that minimizes Eq.(\ref{equation 1}). 

Variance-preserved shot reduction (VPSR) reduces the total measurement cost by introducing a global reduction factor 
\begin{equation}
    \eta = \frac{(\sum_{j=1}^{m} \sigma_{j}(\vec{\theta}))^{2}}{m \sum_{j=1}^{m} \sigma_{j}^{2}(\vec{\theta})},
\end{equation}
which rescales the measurement budget while preserving the target estimator variance. The corresponding shot allocation is 
\begin{equation}
    N_{i} = N \eta.
\end{equation}
Since the Cauchy-Schwarz inequality guarantees $0 \leq \eta \leq 1$, VPSR always assigns less than or equal to uniform allocation \cite{zhu_optimizing_2024}.

In practice, however, the exact variances are unknown and must be estimated from a finite number of pilot measurements. Replacing the true variances by empirical estimates $\vec{\hat{\sigma}} = (\hat{\sigma}_{1}, \hat{\sigma}_{2}, ..., \hat{\sigma}_{m})$ transforms the reduction factor into the nonlinear estimator
\begin{equation}
    \hat{\eta} = g(\vec{\hat{\sigma}}).
    \label{equation 4}
\end{equation}
Because $g(\vec{\hat{\sigma}})$ is nonlinear, $E[g(\vec{\hat{\sigma}})] \neq g(E[\vec{\hat{\sigma}}])$, finite sampling introduces an intrinsic statistical bias. The resulting error propagates directly into the shot allocation, causing systematic under- or over-allocation of measurements. Since VQE optimization depends on accurate energy gradients, these allocation errors accumulate throughout the optimization and frequently produce prolonged optimization plateaus under limited measurement budgets.

To remove this bias, we employ the multivariate delta method. Expanding Eq. (\ref{equation 4}) around the true variance vector gives
\begin{equation}
    g(\vec{\hat{\sigma}}) \approx g(\vec{\sigma}) + \vec{\nabla} g(\vec{\sigma})^{T} (\vec{\hat{\sigma}} - \vec{\sigma}) + \frac{1}{2} (\vec{\hat{\sigma}} - \vec{\sigma})^{T} H (\vec{\hat{\sigma}} - \vec{\sigma}) + ...,
\end{equation}
where $H$ is the Hessian matrix of the second partial derivatives.

As the bias is the difference between the expected value of the estimator with the true value $E[\vec{\hat{\sigma}}] - g(\vec{\sigma})$ \cite{devore_modern_2021}, taking the expectation value yields the expression for bias
\begin{equation}
    \text{Bias}(\hat{\eta}) \approx \vec{\nabla} g(\vec{\sigma})^{T} E[\vec{\hat{\sigma}} - \vec{\sigma}] + \frac{1}{2} E[(\vec{\hat{\sigma}} - \vec{\sigma})^{T} H (\vec{\hat{\sigma}} - \vec{\sigma})].
\end{equation}
Because the reduction factor is a homogeneous function of degree zero, Euler’s theorem ensures that the first-order contribution vanishes identically \cite{lewis_homogeneous_1969}, leaving the leading correction determined entirely by the Hessian
\begin{equation}
    \text{Bias}^{(2)}(\hat{\eta}) = \frac{1}{2} \text{Tr}(H \Sigma),
\end{equation}
where $\Sigma$ is the covariance matrix of the estimated variances.

Independent measurements of different Hamiltonian cliques make $\Sigma$ diagonal, allowing the second-order correction
\begin{equation}
    \text{Bias}^{(2)}(\hat{\eta}) = \frac{1}{2} \sum_{i=1}^{m} H_{ii} \text{Var}(\hat{\sigma}_{i}) = \frac{1}{2} \sum_{i=1}^{m} \frac{\partial^{2} g}{\partial \sigma_{i}^{2}} \text{Var}(\hat{\sigma}_{i}).
\end{equation}
Assuming normally distributed measurement statistics, the variance of the empirical standard deviation obtained from k pilot shots is
\begin{equation}
    \text{Var}(\hat{\sigma}_{i}) \approx \left(\frac{1}{2 \sigma_{i}} \right)^{2} \left( \frac{2 \sigma_{i}^{4}}{k-1} \right),
\end{equation}
leading to the explicit second-order correction
\begin{equation}
    \text{Bias}^{(2)}(\hat{\eta}) \approx \frac{1}{4(k-1)} \sum_{i=1}^{m} \sigma_{i}^{2} \frac{\partial^{2} g}{\partial \sigma_{i}^{2}}.
\end{equation}

The same procedure can be extended systematically to higher orders. Third- and fourth-order expansions introduce corrections arising from the higher-order derivatives of the reduction factor together with the corresponding higher statistical moments of the empirical variance estimator. Their complete analytical expressions are presented in the Supplemental Material. The resulting delta-corrected shot reduction (DCSR) allocates measurements up to fourth-order according to
\begin{equation}
    N_{i} = (\eta - \text{Bias}^{(2)} - \text{Bias}^{(3)} - \text{Bias}^{(4)}) N,
\end{equation}
where
\begin{equation}
    \text{Bias}^{(3)}(\hat{\eta}) \approx \frac{1}{6(k-1)^{2}} \sum_{i=1}^{m} \sigma_{i}^{3} \frac{\partial^{3} g}{\partial \sigma_{i}^{3}},
\end{equation}
and
\begin{equation}
\begin{split}
    \text{Bias}^{(4)}(\hat{\eta}) &\approx \frac{1}{24} \sum_{i=1}^{m} \frac{\partial^{4} g}{\partial \sigma_{i}^{4}} \frac{3 \sigma_{i}^{4} (k+3)}{4(k-1)^{3}} \\
    &\quad + \frac{3}{24} \sum_{i=1}^{m} \sum_{j \neq 1} \frac{\partial^{4} g}{\partial \sigma_{i}^{2} \partial \sigma_{j}^{2}} \frac{\sigma_{i}^{2} \sigma_{j}^{2}}{4(k-1)^{2}}.
\end{split}
\end{equation}


\section{Results and discussion}
To evaluate the performance of the proposed framework, we performed noisy VQE simulations using the Qiskit Aer simulator. The benchmark system was the five-term qubit Hamiltonian of the hydrogen molecule ($E_{FCI} = -0.9752 \text{ Ha}$) and the 27-term, 4-qubit tapered Hamiltonian lithium hydride ($E_{FCI} = -7.863 \text{ Ha}$). Quantum hardware imperfections were modeled using Pauli bit-flip and phase-flip channels acting on state preparation, gate operations, and measurements. To systematically test the limitations of the framework, we performed the $\text{H}_{2}$ simulations with fixed error rate, $p=0.01\%$, across moderate ($N=1200, k=50$) and severe ($N=500, k=25$) finite-sampling regimes to isolate pilot-shot stability, while the LiH simulations utilized a fixed resource budget ($N=18000, k=100$) with error rates $p=0.01\%$ and $p=0.05\%$ to establish the hardware noise threshold. All simulations employed the ADAM optimizer using central finite-difference with identical hyperparameters and were averaged over ten independent realizations.

We first examine the framework under moderate finite-sampling conditions ($N=1200,k=50$) at a physical error rate of $p=0.01\%$. Under these conditions, the finite-sample bias inherent to VPSR can be clearly observed. As shown in Fig. 1(a), the original VPSR algorithm suffers from gradient starvation, manifesting as a transient stall at around iteration 60 and a prolonged optimization plateau hovering just above chemical accuracy from iteration 90 to 160.

\begin{figure*}[tb]
  \centering
  
  \begin{subfigure}[b]{0.48\textwidth}
    \centering
    \includegraphics[width=\textwidth]{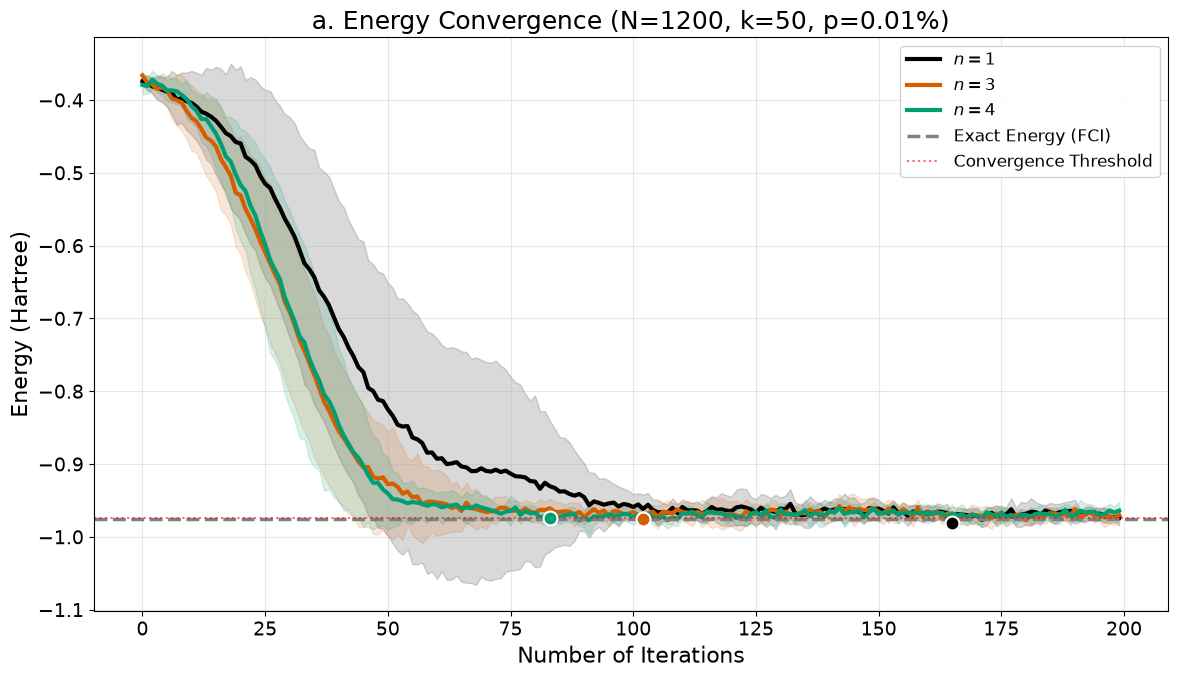} 
    \caption{}
    \label{fig:H2 energy, moderate}
  \end{subfigure}
  \hfill 
  \begin{subfigure}[b]{0.48\textwidth}
    \centering
    \includegraphics[width=\textwidth]{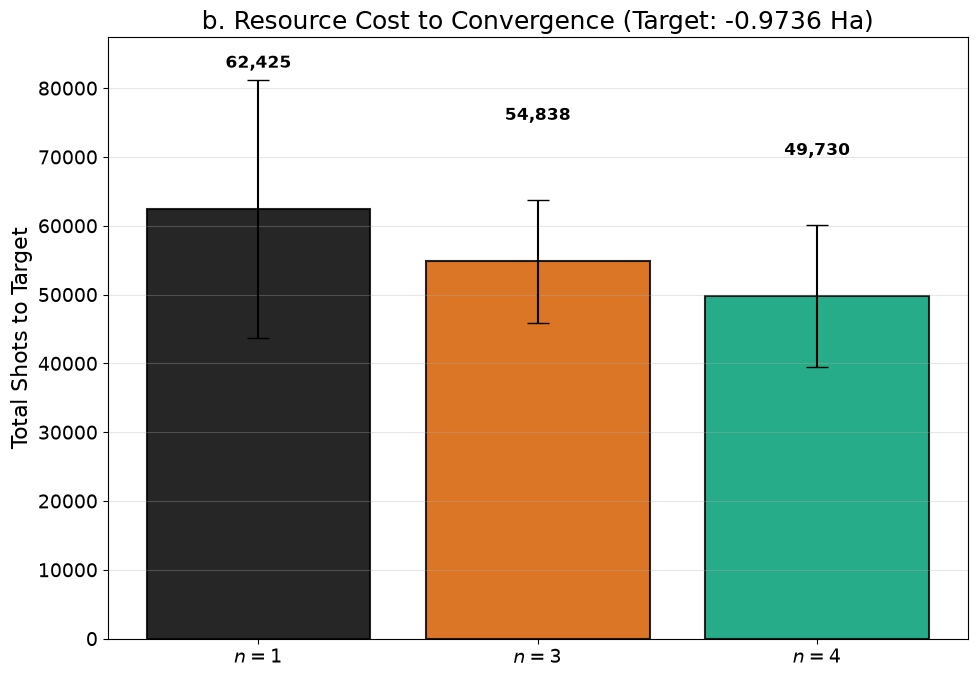} 
    \caption{}
    \label{fig:H2 resources, moderate}
  \end{subfigure}

  \caption{\justifying Performance of DCSR for $\text{H}_2$ under moderate finite-sampling conditions ($N=1200,k=50$) with a Pauli error rate of $0.01\%$. (a) Energy convergence of the VQE optimization (curves truncated at 200 iterations for clarity). Standard VPSR exhibits a transient stall at around iteration 60 and a prolonged optimization plateau above the convergence threshold from iteration 90 to 160.  Higher-order DCSR smoothly bypasses these traps, with the fourth-order exhibiting the fastest convergence. (b) Cumulative measurement cost required to reach chemical accuracy (-0.9736 Ha). Higher-order corrections systematically reduce the total number of measurements, with the fourth-order DCSR requiring 12,965 shots compared to the uncorrected VPSR baseline.}
  \label{fig:H2 Moderate Resources}
\end{figure*}

In contrast, DCSR restores stable optimization by analytically correcting the bias in the empirical reduction factor before shot allocation is performed. All corrected estimators converge substantially faster than the original algorithm, with the fourth-order correction converging almost 100 iterations earlier than VPSR. Rather than producing noisy shot allocations from fluctuating pilot estimates, DCSR accurately tracks the underlying variance landscape, supplying the optimizer with significantly more reliable gradient information. Consequently, optimization proceeds smoothly without the prolonged stalls characteristic of the uncorrected estimator. 

The improvement is equally evident in the cumulative measurement cost shown in Fig. 1(b). Although DCSR introduces additional analytical corrections, it requires no additional quantum sample information. Instead, the improved allocation strategy prevents wasted optimization iterations caused by inaccurate gradients. As a result, higher-order corrections require fewer cumulative measurements to reach the target energy than standard VPSR. Under this condition, the fourth-order correction reduces the total measurement cost by 12,695 shots over VPSR while simultaneously achieving the fastest convergence. 

To thoroughly test the framework’s limitation, we further restricted the resource budget to a severe finite-sample condition ($N=600,k=25$), while maintaining the same error rate. As shown in Fig. 2(a), halving the global shot budget exacerbates the finite-sample bias of VPSR, causing a prolonged optimization plateau that delays convergence until around iteration 190. In contrast, higher-order DCSR completely bypasses this plateau, with the fourth-order achieving convergence near iteration 90, 100 iterations earlier than baseline VPSR. Fig. 2(b) demonstrates that this optimization dominance yields significantly reduced cumulative measurement costs, with the fourth-order estimator saving 5,409 shots compared to the VPSR baseline. 

Additionally, to evaluate the sensitivity of the estimator to small pilot samples sizes, a Monte-Carlo stability analysis simulating NISQ-like conditions revealed a specific breakdown threshold for higher-order terms. At $k \geq 40$, the standard deviations of the second-, third-, and fourth-order corrections are nearly identical. However, at $k \leq 19$, the fourth-order term acts as a noise amplifier as its standard deviation exceeds that of the third-order term. Therefore, below this threshold, the fourth-order correction may be safely discarded. The second-order term generally exhibits the highest standard deviation due to mathematical over-correction, requiring compensation from higher order terms. Notably, while VPSR maintains the lowest standard deviation, its permanent intrinsic bias frequently stalls convergence. The Adam optimizer is highly capable of smoothing the transient volatility of higher-order DCSR terms, but it often struggles to overcome the systemic bias inherent to VPSR.

\begin{figure*}[tb]
  \centering
  
  \begin{subfigure}[b]{0.48\textwidth}
    \centering
    \includegraphics[width=\textwidth]{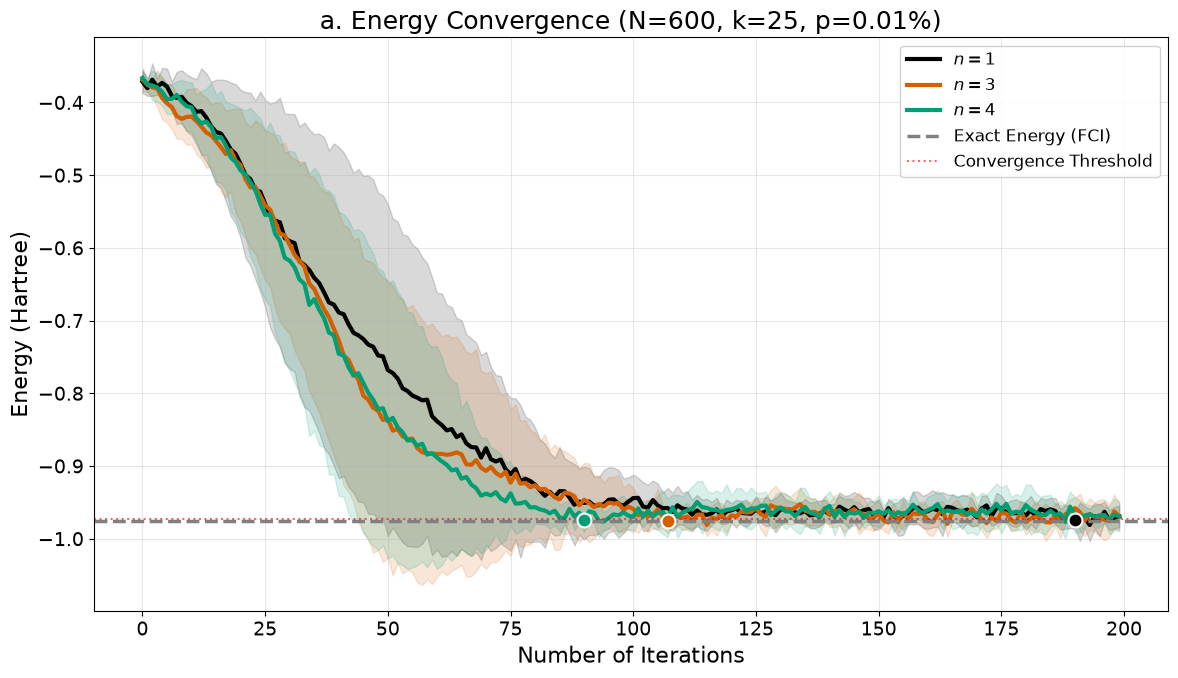} 
    \caption{}
    \label{fig:H2 energy, severe}
  \end{subfigure}
  \hfill 
  \begin{subfigure}[b]{0.48\textwidth}
    \centering
    \includegraphics[width=\textwidth]{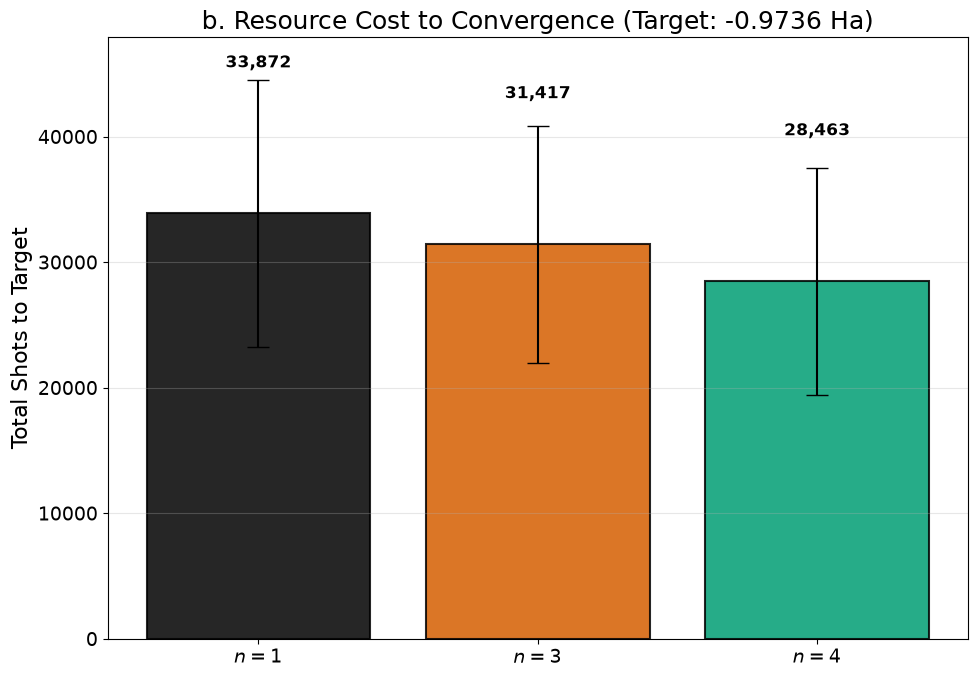}
    \caption{}
    \label{fig:H2 resources, severe}
  \end{subfigure}

  \caption{\justifying Performance of DCSR for $\text{H}_{2}$ under severe finite-sampling condition ($N=600,k=25$) with a Pauli error rate of $0.01\%$. (a) Energy convergence during VQE optimization (curves truncated at 200 iterations for clarity). VPSR experiences a prolonged optimization plateau until iteration 190 due to gradient starvation exacerbated by severe finite-sample conditions. DCSR manages to bypass this trap and achieve convergence much faster than VPSR, with the fourth-order emerging as the dominant trajectory. (b) Cumulative measurements required to reach chemical accuracy (-0.9736 Ha). Increasing the correction order simultaneously accelerates convergence and decreases the total measurement cost, with the fourth-order having the lowest measurement cost.}
  \label{fig:H2 Severe Resources}
\end{figure*}

To establish the framework’s scalability to more complex systems, we next evaluate the 27-term Hamiltonian under hardware noise of $0.01\%$ and an abundant measurement budget ($N=18000,k=100$). As shown in Fig. 3(a), the increased complexity of the Hamiltonian severely degrades the performance of the original VPSR algorithm. The resulting optimization trajectory suffers a pronounced loss of momentum, developing a broad plateau that prevents it from fully converging in the 500 iterations given. Because the analytical corrections compensate directly for the finite-sample bias responsible for inaccurate shot allocation, DCSR systematically suppresses this stagnation. The fourth-order correction provides the most rapid convergence, completely bypassing the region where VPSR becomes trapped. Figure 3(b) demonstrates that this accelerated descent requires strictly fewer measurements as the improved allocation strategy utilizes the physical resources more efficiently to prevent wasted iterations. Consequently, the fourth-order estimator requires only 776,281 shots to achieve chemical accuracy. Compared to the 1.15 million shots expended by the stalled VPSR baseline before it hit the iteration ceiling, the fourth-order DCSR guarantees chemical accuracy while simultaneously reducing the measurement burden by over 380,000 shots. 

\begin{figure*}[tb]
  \centering
  
  \begin{subfigure}[b]{0.48\textwidth}
    \centering
    \includegraphics[width=\textwidth]{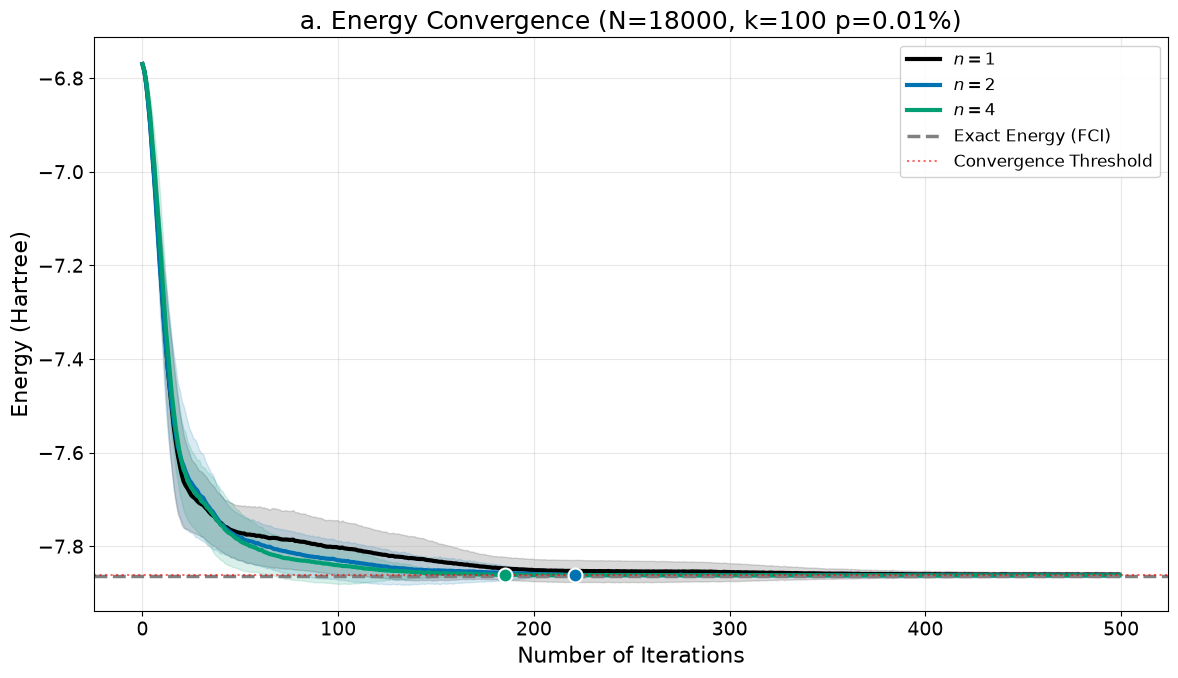} 
    \caption{}
    \label{fig:LiH energy, low noise}
  \end{subfigure}
  \hfill 
  \begin{subfigure}[b]{0.48\textwidth}
    \centering
    \includegraphics[width=\textwidth]{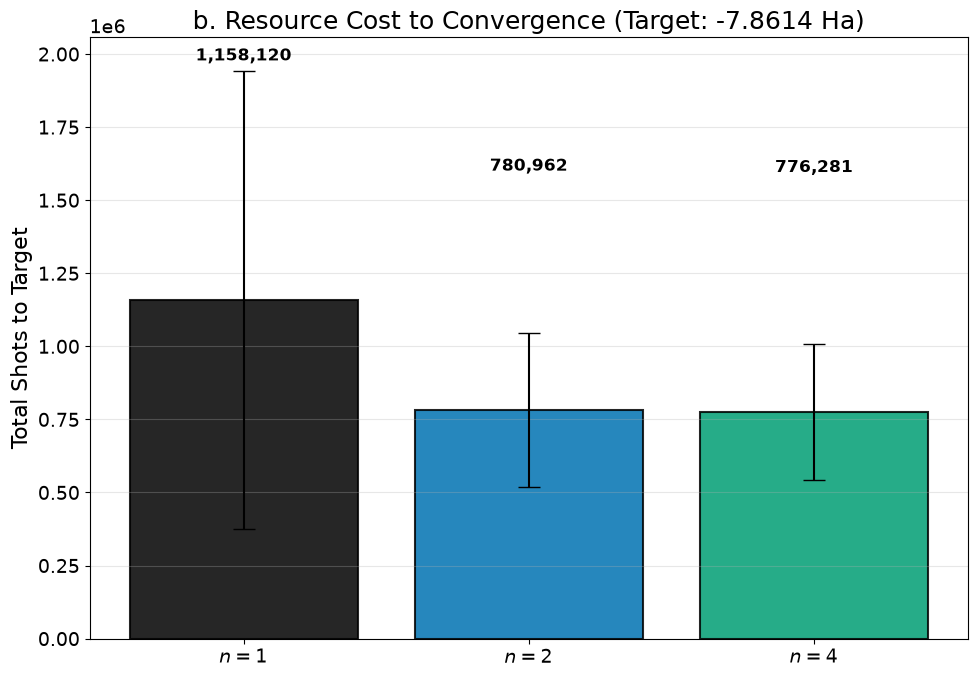}
    \caption{}
    \label{fig:LiH resources, low noise}
  \end{subfigure}

  \caption{\justifying Performance of DCSR for LiH with shot budget $N=18000,k=100$, and with a Pauli error rate of $0.01\%$. (a) Energy convergence during VQE optimization (third-order omitted for clarity).  Increased Hamiltonian complexity induces a severe optimization plateau in baseline VPSR.  In contrast, DCSR successfully bypasses this stagnation, with the fourth-order correction reaching the target energy near iteration 185. (b) Cumulative measurement cost required to reach chemical accuracy (-7.8614 Ha). The fourth-order estimator reduces the total measurement burden by 381,839 shots compared to the stalled baseline.}
  \label{fig:LiH Low Noises}
\end{figure*}

Finally, to evaluate the framework's resilience across varying QPU hardware architectures, a sensitivity analysis was conducted across a spectrum of physical gate error rates. Empirical simulations identify $p=0.05\%$ as the critical noise breakdown threshold for achieving chemical accuracy. At this specific error rate, the corrected trajectories asymptotically converge precisely at the chemical accuracy limit. Under these extreme conditions, the fourth-order correction behaves as a noise-amplifier and is therefore excluded to minimize visual clutter. 

Fig. 4(a) demonstrates this effect, where baseline VPSR and second-order DCSR trajectories plateau definitively above the convergence threshold. In contrast, third-order DCSR successfully convergences to chemical accuracy near iteration 410. This verifies that third-order DCSR remains the only viable strategy at this limit, achieving chemical accuracy while reducing the cumulative measurement resource by 312,656 shots compared to the accumulated total shots of VPSR over 500 iterations. Beyond this $0.05\%$ threshold, the noise floor entirely overwhelms the gradient, causing the energy evaluations for all correction orders to plateau definitively above the target energy. 

\begin{figure*}[tb]
  \centering
  
  \begin{subfigure}[b]{0.48\textwidth}
    \centering
    \includegraphics[width=\textwidth]{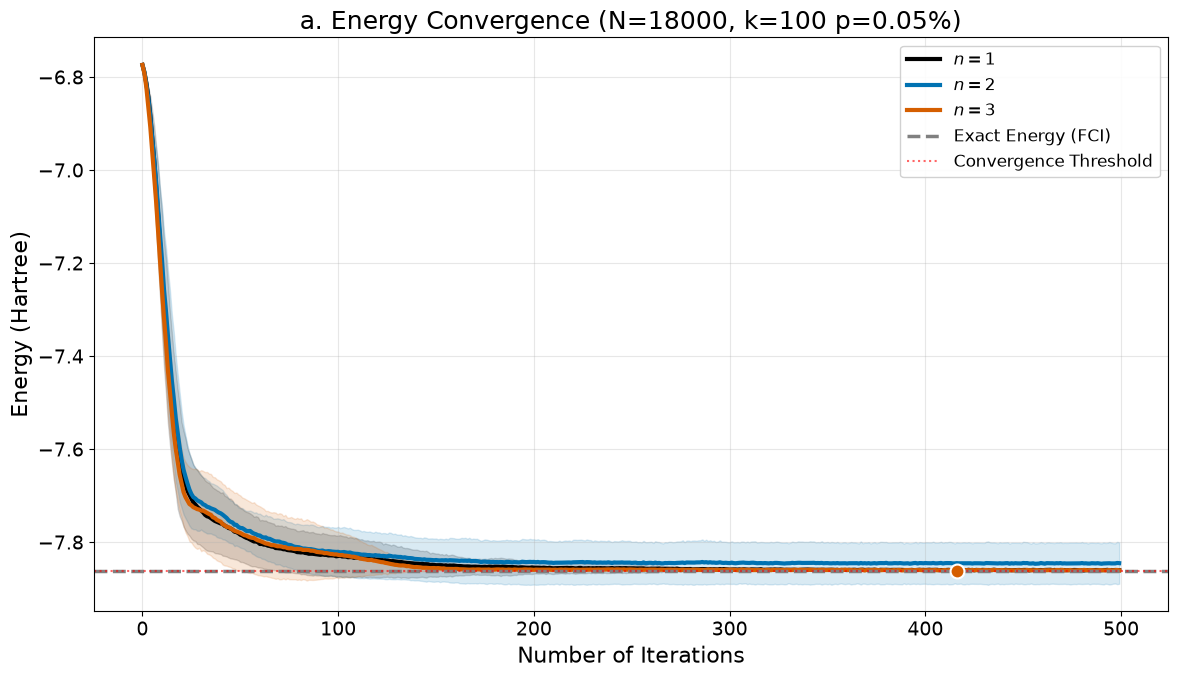} 
    \caption{}
    \label{fig:LiH energy, high noise}
  \end{subfigure}
  \hfill 
  \begin{subfigure}[b]{0.48\textwidth}
    \centering
    \includegraphics[width=\textwidth]{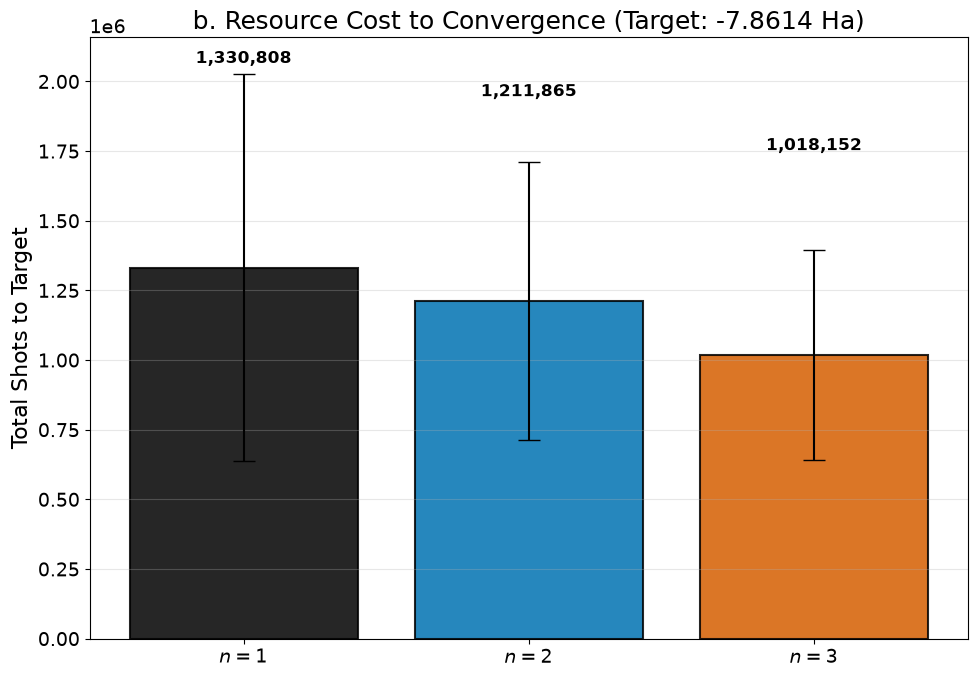}
    \caption{}
    \label{fig:LiH resources, high noise}
  \end{subfigure}

  \caption{\justifying Performance of DCSR for LiH with shot budget of $N=18000,k=100$, and with a Pauli error rate of $0.05\%$. (a) Energy convergence during VQE optimization. Under the high noise, the trajectories of baseline VPSR and second-order DCSR plateau above the convergence threshold. However, third-order DCSR successfully converges to chemical accuracy, proving its robustness under this specific restrictive condition.  (b) Cumulative measurement required to reach chemical accuracy (-7.8614 Ha). Increasing the correction order decreases the total measurement cost, with third-order DCSR having the lowest measurement cost.}
  \label{fig:LiH High Noises}
\end{figure*}

Across all operating regimes, a consistent picture emerges. The dominant source of instability in adaptive shot reduction is not merely hardware noise but the finite-sample bias introduced by estimating variances from limited pilot measurements. By analytically removing this bias through the multivariate delta method, DCSR simultaneously improves optimization stability and measurement efficiency. While incorporating higher-order corrections yields progressively more accurate estimates of the optimal shot allocation under moderate conditions, our analysis reveals that the estimator is bounded by physical and statistical noise floors. In severe finite-sampling ($k \leq 20$) or high-noise regimes ($p \geq 0.05\%$), the third-order correction emerges as the robust optimal compromise. Consequently, DCSR transforms adaptive shot reduction from a heuristic sampling strategy into a statistically rigorous measurement-allocation framework suitable for noisy quantum hardware.

\section{Conclusions}
\label{sec:conclusion}
We have introduced delta-corrected shot reduction (DCSR), an analytically derived framework for mitigating the finite-sample bias that limits the performance of adaptive shot-allocation methods in variational quantum eigensolver. By applying the multivariate delta method, we obtained closed-form higher-order corrections to the VPSR reduction factor, enabling statistically unbiased measurement allocation under finite sampling. Unlike heuristic regularization schemes, the proposed approach follows directly from the asymptotic properties of the empirical variance estimator and therefore provides a systematic route to improving adaptive measurement strategies.

Numerical simulations demonstrate that the analytical corrections substantially improve optimization stability across both noise-limited and resource-limited regimes. By removing the bias responsible for inaccurate shot allocation, DCSR suppresses the optimization plateaus frequently observed in the original VPSR algorithm while preserving its measurement efficiency. In the most demanding operating conditions, higher-order corrections not only accelerate convergence but also reduce the cumulative number of measurements required to reach the target energy, showing that improved statistical accuracy and measurement efficiency can be achieved simultaneously.

This accelerated convergence is achieved with negligible classical computational overhead, remaining under $100 \text{ }\mu \text{s}$ even when calculating complex fourth-order cross terms. Because DCSR extracts the same variance information from the pilot shots as with VPSR, our framework performs with greater effectiveness and efficiency without requiring additional sample information nor incurring meaningful classical penalties.

Furthermore, DCSR is complementary to overlap-based and gradient-based allocation strategies. An overlap-based grouping method is implemented in grouping commuting Hamiltonians into cliques as a baseline step. Gradient-based allocation strategies optimize the parameter update path rather than correcting the finite-sample bias. Consequently, DCSR can complement and be complemented by these methods to achieve greater accuracy.

The analytical derivation of the present framework is unaffected by the specific molecular Hamiltonian, ansatz construction, and classical optimization algorithm. However, while the empirical stability of the higher-order corrections remains coupled to classical optimizer variance and baseline shot budgets, the analytical nature of the derivations themselves makes DCSR broadly adaptable to adaptive measurement protocols beyond standard VQE. More generally, these results demonstrate that finite-sample effects constitute an important source of optimization error in variational quantum algorithms and that analytical bias correction offers an effective strategy for mitigating this limitation. As quantum hardware continues to scale while remaining measurement constrained, statistically rigorous resource-allocation methods such as DCSR becomes an important component of practical quantum simulation and other variational quantum computing applications.

\begin{acknowledgments}
This work was supported by grants from National Research and Innovation Agency (BRIN), Indonesia and Centre for Quantum Technologies (CQT), National University of Singapore, Singapore. 
\end{acknowledgments}
\section*{Data Availability}
\noindent The source code for implementing the DCSR framework up to fourth-order to estimate the ground state energy of the $\text{H}_2$ and LiH molecule along with the raw data for Figs. 1(a) and 1(b) are available in the online repository: \url{https://github.com/JamieEGhozaly/DCSR-VQE-Framework-and-Data/tree/main}
\section*{Appendix A: First-Order Delta Correction }
To determine the finite-sample bias of the VPSR reduction factor, we expand $g(\vec{\hat{\sigma}})$ about the true standard-deviation vector $\vec{\sigma}$ to second order, 
\begin{equation}
    g(\vec{\hat{\sigma}}) \approx g(\vec{\sigma}) + \vec{\nabla} g(\vec{\sigma})^{T} (\vec{\hat{\sigma}} - \vec{\sigma}) + \frac{1}{2} (\vec{\hat{\sigma}} - \vec{\sigma})^{T} H (\vec{\hat{\sigma}} - \vec{\sigma}),
\end{equation}
where $H$ is the Hessian matrix of $g$. Taking the expectation value and subtracting $g(\vec{\hat{\sigma}})$ gives
\begin{equation}
    \text{Bias}(\hat{\eta}) \approx \vec{\nabla} g(\vec{\sigma})^{T} E[\vec{\hat{\sigma}} - \vec{\sigma}] + \frac{1}{2} E[(\vec{\hat{\sigma}} - \vec{\sigma})^{T} H (\vec{\hat{\sigma}} - \vec{\sigma})].
\end{equation}
The first-order contribution is determined by the finite-sample bias of the sample standard deviation. For $k$ samples, its expectation can be written as
\begin{equation}
    E[\hat{\sigma_i}] = c_{k} \sigma_{i},
\end{equation}
where $c_k$ depends on the sample size. Hence, 
\begin{equation}
    E[\hat{\sigma}_{i} - \sigma_i] = (c_{k}-1) \sigma_{i},
\end{equation}
and the first order contribution becomes,
\begin{equation}
    (c_{k}-1)\sum_{i=1}^{m} \sigma_{i} \frac{\partial g}{\partial \sigma_{i}}.
    \label{equation 18}
\end{equation}
The VPSR reduction factor is homogeneous of degree zero. Euler’s theorem for homogeneous functions, 
\begin{equation}
    \sum_{i=1}^{m} x_{i} \frac{\partial f(x_{i})}{\partial x_{i}} = n \cdot f(\vec{x}),
\end{equation}
\cite{lewis_homogeneous_1969} therefore gives,
\begin{equation}
    \sum_{i=1}^{m} \sigma_{i} \frac{\partial g(\sigma_{i})}{\partial \sigma_{i}} = n \cdot g(\vec{\hat{\sigma}}) = 0 \cdot g(\vec{\hat{\sigma}}) = 0.
\end{equation}
Consequently, the first-order contribution in Eq. (\ref{equation 18}) vanishes identically,
\begin{equation}
    \text{Bias}^{(1)} (\hat{\eta}) = 0.
\end{equation}
Thus, although the individual sample standard deviations are biased at finite $k$, their common multiplicative bias does not propagate to the VPSR reduction factor at first order. The leading finite-sample contribution to the bias arises from the second-order term.

\section*{APPENDIX B: Second-Order Delta Correction}
The second-order contribution to the finite-sample bias follows from the second-order multivariate Taylor expansion 
\begin{equation}
    g(\vec{\hat{\sigma}}) \approx g(\vec{\sigma}) + \vec{\nabla} g(\vec{\sigma})^{T} (\vec{\hat{\sigma}} - \vec{\sigma}) + \frac{1}{2} (\vec{\hat{\sigma}} - \vec{\sigma})^{T} H (\vec{\hat{\sigma}} - \vec{\sigma}),
\end{equation}
where $H$ denotes the Hessian of $g(\vec{\hat{\sigma}})$. Taking the expectation value and using the vanishing of the first-order contribution gives
\begin{equation}
    \text{Bias}^{(2)} (\hat{\eta}) = E[g(\vec{\hat{\sigma}})] - g(\vec{\sigma}) \approx \frac{1}{2} E[(\vec{\hat{\sigma}} - \vec{\sigma})^{T}H(\vec{\hat{\sigma}} - \vec{\sigma})].
\end{equation}
Writing the quadratic form as a trace and using the linearity of the trace and expectation operators,
\begin{equation}
    \text{Bias}^{(2)} (\hat{\eta}) \approx \frac{1}{2} \text{Tr}(H E[(\vec{\hat{\sigma}} - \vec{\sigma})(\vec{\hat{\sigma}} - \vec{\sigma})^{T}]) = \frac{1}{2} \text{Tr}(H \Sigma),
\end{equation}
where $\Sigma$ is the covariance matrix of $\vec{\hat{\sigma}}$. Since measurements associated with distinct cliques are independent, $\Sigma$ is diagonal, and therefore
\begin{equation}
    \text{Bias}^{(2)} (\hat{\eta}) = \frac{1}{2} \sum_{i=1}^{m} H_{ii} \text{Var}(\hat{\sigma}_{i}) =\frac{1}{2} \sum_{i=1}^{m} \frac{\partial^{2} g}{\partial \sigma_{i}^{2}} \text{Var}(\hat{\sigma}_{i}).
    \label{equation 25}
\end{equation}
For the VPSR reduction factor, defining $S_{1} = \sum_{j=1}^{m} \sigma_{j}$, $S_{2} = \sum_{j=1}^{m} \sigma_{j}^{2}$, we get $g(\vec{\sigma}) = \frac{S_{1}^{2}}{m S_{2}}$ and its first derivative with respect to $\sigma_{i}$ is 
\begin{equation}
    \frac{\partial g}{\partial \sigma_{i}} = \frac{2 S_{1} S_{2} - 2 S_{1}^{2} \sigma_{i}}{m S_{2}^{2}} = \frac{2 S_{1}}{m S_{2}} - \frac{2 S_{1}^{2}}{m S_{2}^{2}} \sigma_{i},
\end{equation}
and the corresponding Hessian element is
\begin{equation}
    \frac{\partial^{2} g}{\partial \sigma_{i}^{2}} = \frac{2}{m S_{2}} - \frac{8 S_{1} \sigma_{i}}{m S_{2}^{2}} - \frac{2 S_{1}^{2}}{m S_{2}^{2}} + \frac{8 S_{1}^{2} \sigma_{i}^{2}}{m S_{2}^{3}}.
    \label{equation 27}
\end{equation}
The variance of each sample standard deviation is obtained from the sample variance. For $k$ independent pilot shots 
\begin{equation}
    S_{i}^{2} = \frac{1}{k-1} \sum_{s=1}^{k} (e_{i}^{s}(\theta) - \bar{E}_{i}(\theta))^{2},
\end{equation}
and, under the Gaussian measurement-error approximation
\begin{equation}
    \frac{(k-1) S_{i}^{2}}{\sigma_{i}^{2}} \sim \chi_{k-1}^{2}.
    \label{equation 29}
\end{equation}
Since a $\chi_{\nu}^{2}$ random variable has variance $2 \nu$ \cite{cramer_mathematical_1999}, by applying standard variance scaling properties $\text{Var}(cX) = c^{2} \text{Var}(X)$ \cite{nielsen_quantum_2012}, we can isolate the variance of $\text{Var}(S_{i}^{2})$:
\begin{equation}
    \text{Var}(S_{i}^{2}) = \left( \frac{\sigma_{i}^{2}}{k-1} \right)^2 2(k-1) = \frac{2 \sigma_{i}^{4}}{k-1}.
\end{equation}
We require the variance of $\hat{\sigma}_{i} =\sqrt{S_{i}^{2}}$. Applying the first-order univariate delta method to $f(x) = \sqrt{x}$, for which $f'(x) = \frac{1}{2 \sqrt{x}}$ gives
\begin{equation}
    \text{Var}(\hat{\sigma}_{i}) \approx [f'(\sigma_{i}^{2})]^2 \text{Var}(S_{i}^{2}) = \left( \frac{1}{2 \sigma_{i}} \right)^{2} \left( \frac{2 \sigma_{i}^{4}}{k-1} \right) = \frac{\sigma_{i}^{2}}{2 (k-1)}.
    \label{equation 31}
\end{equation}
Substitution into Eq. (\ref{equation 25}) yields the second-order finite-sample bias,
\begin{equation}
    \text{Bias}^{(2)}(\hat{\eta}) \approx \frac{1}{4(k-1)} \sum_{i=1}^{m} \sigma_{i}^{2} \frac{\partial^{2} g}{\partial \sigma_{i}^{2}}.
\end{equation}
Using Eq. (\ref{equation 27}), this correction is completely determined by the clique standard deviations and the corresponding second derivatives of the reduction factor. The second-order contribution therefore scales as $\mathcal{O}[(k-1)^{-1}]$.

\section*{APPENDIX C: Third-Order Delta Correction}
The third-order contribution to the multivariate delta expansion is
\begin{equation}
\begin{aligned}
    \text{Bias}^{(3)}\left(\hat{\eta}\right) = 
    & \frac{1}{6} E \Bigg[ \sum_{i=1}^{m} \sum_{j=1}^{m} \sum_{k=1}^{m} \frac{\partial^3g}{\partial\sigma_i \partial\sigma_j \partial\sigma_k} \\
    & \times \left(\hat{\sigma}_i - \sigma_i\right) \left(\hat{\sigma}_j - \sigma_j\right) \left(\hat{\sigma}_k - \sigma_k\right) \Bigg].
\end{aligned}
\end{equation}
Because the measurement samples associated with distinct cliques are independent, mixed central moments factorize. Moreover, the sample standard deviation is asymptotically unbiased, so that 
\begin{equation}
    E[(\hat{\sigma}_{i} - \sigma_{i})] \rightarrow 0, (k \rightarrow \infty)
\end{equation}
Consequently, the leading third-order contribution is obtained from the diagonal terms $i=j=k$, for which the third central moment does not factorize into the cube of the first central moment
\begin{equation}
    E[(\hat{\sigma}_{i} - \sigma_{i})^{3}] \neq (E[\hat{\sigma}_{i} - \sigma_{i}])^{3}.
\end{equation}
Therefore, the third-order bias is expressed by
\begin{equation}
    \text{Bias}^{(3)}(\hat{\eta}) = \frac{1}{6} \sum_{i=1}^{m} \frac{\partial^{3} g}{\partial \sigma_{i}^{3}} E[(\hat{\sigma}_{i} - \sigma_{i})^{3}].
\end{equation}
Using Eq. (\ref{equation 27}), the required third derivative of the VPSR reduction factor is 
\begin{equation}
    \frac{\partial^3g}{\partial\sigma_i^3}=\frac{12\left(2\sigma_iS_1S_2\left(2\sigma_i+S_1\right)-S_2^2\left(\sigma_i+S_1\right)-4\sigma_i^3S_1^2\right)}{mS_2^4}.
    \label{equation 37}
\end{equation}
We next evaluate the third central moment of $\hat{\sigma}_{i}$. From Eq. (\ref{equation 29}), a chi-square random variable with $\nu$ degrees of freedom has third central moment $8 \nu$ \cite{cramer_mathematical_1999}. Therefore,
\begin{equation}
    E[(S_{i}^{2} - \sigma_{i}^{2})^{3}] = \left( \frac{\sigma_{i}^{2}}{k-1} \right)^{3} 8 (k-1) = \frac{8 \sigma_{i}^{6}}{(k-1)^{2}}.
\end{equation}
To propagate this moment from $S_{i}^{2}$ to $\hat{\sigma}_{i} = \sqrt{S_{i}^{2}}$, we use the first-order univariate delta method with 
\begin{equation}
    f(x) = \sqrt{x}, \qquad f'(\sigma_i^2) = \frac{1}{2\sigma_i}.
\end{equation}
Hence,
\begin{gather}
    \hat{\sigma}_i - \sigma_i \approx f'(\sigma_i^2)(S_i^2 - \sigma_i^2), \notag \\
    (\hat{\sigma}_i - \sigma_i)^3 \approx [f'(\sigma_i^2)(S_i^2 - \sigma_i^2)]^3, \label{eq:delta_method_cubed} \\
    (\hat{\sigma}_i - \sigma_i)^3 \approx [f'(\sigma_i^2)]^3 (S_i^2 - \sigma_i^2)^3. \notag
\end{gather}
and, to leading order, 
\begin{align}
    E[(\hat{\sigma}_i - \sigma_i)^3] &\approx [f'(\sigma_i^2)]^3 E[(S_i^2 - \sigma_i^2)^3] \notag \\
    E[(\hat{\sigma}_i - \sigma_i)^3] &\approx \left( \frac{1}{2\sigma_i} \right) \left( \frac{8\sigma_i^6}{(k-1)^2} \right) \label{eq:third_moment} \\
    &= \left( \frac{1}{8\sigma_i^3} \right) \left( \frac{8\sigma_i^6}{(k-1)^2} \right) = \frac{\sigma_i^3}{(k-1)^2}. \notag
\end{align}
Substitution into Eq. (\ref{equation 37}) gives the third-order bias correction,
\begin{align}
    \text{Bias}^{(3)}(\hat{\eta}) &\approx \frac{1}{6} \sum_{i=1}^{m}\frac{\partial^{3}g}{\partial \sigma_{i}^{3}} E[(\hat{\sigma}_{i} - \sigma_{i})^{3}] \\
    &= \frac{1}{6(k-1)^{2}} \sum_{i=1}^{m}\sigma_{i}^{3} \frac{\partial^{3} g}{\partial \sigma_{i}^{3}} \notag
\end{align}

\section*{APPENDIX D: Fourth-Order Delta Correction}
The fourth-order contribution to the multivariate delta expansion is
\begin{equation}
    \text{Bias}^{(4)}\left(\hat{\eta}\right) = \frac{1}{24} E \left[\sum_{i,j,k,l=1}^{m} \frac{\partial^4g}{\partial\sigma_i \partial\sigma_j \partial\sigma_k \partial\sigma_l} \times D_{i} D_{j} D_{k} D_{l} \right],
\end{equation}
where
\begin{equation}
    D_{i} = \hat{\sigma}_{i} - \sigma_{i}.
\end{equation}
Because the clique estimators are independent, the mixed moments factorize whenever they contain independent indices. Terms with four distinct indices, three distinct indices, or one repeated index and two distinct indices vanish because the corresponding first central moments are zero. The surviving contributions are therefore of two types $i=j=k=$l, and pairwise contractions $i=j,k=l,i \neq k$. Specifically,
\begin{equation}
    E[D_{i}^{4}] = E[(\hat{\sigma}_{i} - \sigma_{i})^{4}],
\end{equation}
while
\begin{equation}
    E[D_{i}^{2} D_{j}^{2}] = E[D_{i}^{2}] E[D_{j}^{2}] = \text{Var}(\hat{\sigma}_{i}) \text{Var}(\hat{\sigma}_{j}), i \neq j.
\end{equation}
There are $\frac{\binom{4}{2}}{2} = 3$ distinct pairwise contractions, thus the fourth-order bias expression becomes
\begin{equation}
\begin{split}
    \text{Bias}^{(4)}(\hat{\eta}) &= \frac{1}{24} \sum_{i=1}^{m} \frac{\partial^4 g}{\partial \sigma_i^4} E[(\hat{\sigma}_i - \sigma_i)^4] \\
    &\quad + \frac{3}{24} \sum_{i=1}^{m} \sum_{j \neq i} \frac{\partial^4 g}{\partial \sigma_i^2 \partial \sigma_j^2} \operatorname{Var}(\hat{\sigma}_i) \operatorname{Var}(\hat{\sigma}_j).
\end{split}
\label{equation 47}
\end{equation}
The required fourth derivatives follow from differentiating Eq. (\ref{equation 37}). For the diagonal derivative,
\begin{equation}
    \frac{\partial^{4} g}{\partial \sigma_{i}^{4}} = \frac{U' S_{2} - 8 \sigma_{i} U}{m S_{2}^{5}},
\end{equation}
where
\begin{equation}
    U = 48 \sigma_{i}^{2} S_{1} S_{2} + 24 \sigma_{i}S_{1}^{2} S_{2} - 12 \sigma_{i} S_{2} - 12 S_{1} S_{2}^{2} - 48 \sigma_{i}^{3} S_{1}^{2},
\end{equation}
and
\begin{equation}
    U' = 24(S_{1}^{2} S_{2} - 4 \sigma_{i}^{2} S_{1}^{2} + 4 \sigma_{i} S_{1} S_{2} - S_{2}^{2}).
\end{equation}
For the mixed derivative, differentiation of Eq. (\ref{equation 37}) gives
\begin{equation}
    \frac{\partial^{3} g}{\partial \sigma_{i}^{2} \partial \sigma_{j}} = \frac{W}{m S_{2}^{4}},
\end{equation}
with
\begin{equation}
\begin{split}
    &= 16 \sigma_i^2 S_1 S_2 + 32 \sigma_i \sigma_j S_1 S_2 + 8 \sigma_j S_1^2 S_2 \\
    &\quad - 8 \sigma_i S_2^2 - 4 \sigma_j S_2^2 - 4 S_1 S_2^2 - 48 \sigma_i^2 \sigma_j S_1^2.
\end{split}
\end{equation}
Differentiating once more with respect to ${\sigma }_{j}$ yields
\begin{equation}
    \frac{{\partial }^{4}g}{\partial {\sigma }_{i}^{2}\partial {\sigma }_{j}^{2}}=\frac{W^{\prime }\left(mS_{2}^{4}\right)-W\left(8m{\sigma }_{j}S_{2}^{3}\right)}{m^{2}S_{2}^{8}},
\end{equation}
where
\begin{equation}
\begin{split}
    W' &= 8 \Bigl( S_1^2 S_2 - 6 \sigma_i^2 S_1^2 + 2 \sigma_j^2 S_1^2 + 4 \sigma_i S_1 S_2 \\
    &\qquad - 8 \sigma_i^2 \sigma_j S_1 + 8 \sigma_i \sigma_j^2 S_1 - S_2^2 + 2 \sigma_i^2 S_2 - 2 \sigma_j^2 S_2 \Bigr).
\end{split}
\end{equation}
The variance factors required in Eq. (\ref{equation 47}) follow from Eq. (\ref{equation 31})
\begin{equation}
    \text{Var}\left({\hat{\sigma }}_{i}\right)\approx \frac{{\sigma }_{i}^{2}}{2\left(k-1\right)},
\end{equation}
and hence 
\begin{equation}
    \text{Var}\left({\hat{\sigma }}_{i}\right)\text{Var}\left({\hat{\sigma }}_{j}\right)\approx \frac{{\sigma }_{i}^{2}{\sigma }_{j}^{2}}{4{\left(k-1\right)}^{2}}.
\end{equation}
It remains to determine the fourth central moment of ${\hat{\sigma }}_{i}$. The fourth central moment of a ${\chi }_{\nu }^{2}$ random variable with $\nu$ degrees of freedom is 1$2\nu \left(\nu +4\right)$. Setting $\nu = k-1$ gives
\begin{equation}
\begin{split}
    E[(S_i^2 - \sigma_i^2)^4] &= \left(\frac{\sigma_i^2}{k-1}\right)^{\!4} 12(k-1)(k+3) \\
    &= \frac{12\sigma_i^8 (k+3)}{(k-1)^3}.
\end{split}
\end{equation}
Applying the first-order delta method to $f(x)=\sqrt{x}$,
\begin{equation}
    {\hat{\sigma }}_{i}-{\sigma }_{i}\approx f^{\prime }({\sigma }_{i}^{2})(S_{i}^{2}-{\sigma }_{i}^{2}),\ \ \ \ f^{\prime }({\sigma }_{i}^{2})=\frac{1}{2{\sigma }_{i}},
\end{equation}
gives
\begin{equation}
\begin{split}
    E[(\hat{\sigma}_i - \sigma_i)^4] &\approx [f'(\sigma_i^2)]^4 E[(S_i^2 - \sigma_i^2)^4] \\
    &= \left(\frac{1}{2\sigma_i}\right)^{\!4} \left(\frac{12\sigma_i^8 (k+3)}{(k-1)^3}\right) = \frac{3\sigma_i^4 (k+3)}{4(k-1)^3}.
\end{split}
\end{equation}
The fourth-order correction is
\begin{equation}
\begin{split}
    \operatorname{Bias}^{(4)}(\hat{\eta}) 
    &= \frac{1}{24} \sum_{i=1}^{m} \frac{\partial^4 g}{\partial \sigma_i^4} \frac{3\sigma_i^4(k+3)}{4(k-1)^3} \\
    &\quad + \frac{3}{24} \sum_{i=1}^{m} \sum_{j\neq i} \frac{\partial^4 g}{\partial \sigma_i^2 \partial \sigma_j^2} \frac{\sigma_i^2 \sigma_j^2}{4(k-1)^2} \\
    &= \frac{k+3}{32(k-1)^3} \sum_{i=1}^{m} \sigma_i^4 \frac{\partial^4 g}{\partial \sigma_i^4} \\
    &\quad + \frac{1}{32(k-1)^2} \sum_{i=1}^{m} \sum_{j\neq i} \sigma_i^2 \sigma_j^2 \frac{\partial^4 g}{\partial \sigma_i^2 \partial \sigma_j^2}.
\end{split}
\end{equation}
\newpage

\bibliographystyle{apsrev4-2} 
\bibliography{jamie_references}     

@article{preskill_quantum_2018,
	title = {Quantum {Computing} in the {NISQ} era and beyond},
	volume = {2},
	copyright = {https://creativecommons.org/licenses/by/4.0/},
	issn = {2521-327X},
	url = {https://quantum-journal.org/papers/q-2018-08-06-79/},
	doi = {10.22331/q-2018-08-06-79},
	urldate = {2026-06-12},
	journal = {Quantum},
	author = {Preskill, John},
	month = aug,
	year = {2018},
	pages = {79},
}

@book{nielsen_quantum_2012,
	edition = {1},
	title = {Quantum {Computation} and {Quantum} {Information}: 10th {Anniversary} {Edition}},
	copyright = {https://www.cambridge.org/core/terms},
	isbn = {978-1-107-00217-3 978-0-511-97666-7},
	shorttitle = {Quantum {Computation} and {Quantum} {Information}},
	url = {https://www.cambridge.org/core/product/identifier/9780511976667/type/book},
	doi = {10.1017/CBO9780511976667},
	urldate = {2026-06-12},
	publisher = {Cambridge University Press},
	author = {Nielsen, Michael A. and Chuang, Isaac L.},
	month = jun,
	year = {2012},
}

@book{cramer_mathematical_1999,
	series = {Princeton {Landmarks} in {Mathematics} and {Physics}},
	title = {Mathematical {Methods} of {Statistics}},
	isbn = {978-0-691-00547-8},
	url = {https://books.google.co.id/books?id=jV2YDwAAQBAJ},
	publisher = {Princeton University Press},
	author = {Cramér, H.},
	year = {1999},
	lccn = {lc99012156},
}

@incollection{lewis_homogeneous_1969,
	address = {London},
	title = {Homogeneous {Functions} and {Euler}’s {Theorem}},
	isbn = {978-0-333-01021-1 978-1-349-15324-4},
	url = {http://link.springer.com/10.1007/978-1-349-15324-4_22},
	doi = {10.1007/978-1-349-15324-4_22},
	urldate = {2026-06-12},
	booktitle = {An {Introduction} to {Mathematics}},
	publisher = {Palgrave Macmillan UK},
	author = {Lewis, J. Parry},
	collaborator = {Lewis, J. Parry},
	year = {1969},
	pages = {297--303},
}

@book{devore_modern_2021,
	address = {Cham},
	series = {Springer {Texts} in {Statistics}},
	title = {Modern {Mathematical} {Statistics} with {Applications}},
	copyright = {https://www.springer.com/tdm},
	isbn = {978-3-030-55155-1 978-3-030-55156-8},
	url = {https://link.springer.com/10.1007/978-3-030-55156-8},
	doi = {10.1007/978-3-030-55156-8},
	urldate = {2026-06-12},
	publisher = {Springer International Publishing},
	author = {Devore, Jay L. and Berk, Kenneth N. and Carlton, Matthew A.},
	year = {2021},
}

@article{bharti_noisy_2022,
	title = {Noisy intermediate-scale quantum algorithms},
	volume = {94},
	number = {1},
	journal = {Reviews of Modern Physics},
	publisher = {APS},
	author = {Bharti, Kishor and Cervera-Lierta, Alba and Kyaw, Thi Ha and Haug, Tobias and Alperin-Lea, Sumner and Anand, Abhinav and Degroote, Matthias and Heimonen, Hermanni and Kottmann, Jakob S and Menke, Tim and {others}},
	year = {2022},
	pages = {015004},
}

@article{mcardle_quantum_2020,
	title = {Quantum computational chemistry},
	volume = {92},
	number = {1},
	journal = {Reviews of Modern Physics},
	publisher = {APS},
	author = {McArdle, Sam and Endo, Suguru and Aspuru-Guzik, Alán and Benjamin, Simon C and Yuan, Xiao},
	year = {2020},
	pages = {015003},
}

@article{cao_quantum_2019,
	title = {Quantum chemistry in the age of quantum computing},
	volume = {119},
	number = {19},
	journal = {Chemical reviews},
	publisher = {ACS Publications},
	author = {Cao, Yudong and Romero, Jonathan and Olson, Jonathan P and Degroote, Matthias and Johnson, Peter D and Kieferová, Mária and Kivlichan, Ian D and Menke, Tim and Peropadre, Borja and Sawaya, Nicolas PD and {others}},
	year = {2019},
	pages = {10856--10915},
}

@article{bauer_quantum_2020,
	title = {Quantum algorithms for quantum chemistry and quantum materials science},
	volume = {120},
	number = {22},
	journal = {Chemical reviews},
	publisher = {ACS Publications},
	author = {Bauer, Bela and Bravyi, Sergey and Motta, Mario and Chan, Garnet Kin-Lic},
	year = {2020},
	pages = {12685--12717},
}

@article{lau_nisq_2022,
	title = {{NISQ} computing: where are we and where do we go?},
	volume = {32},
	number = {1},
	journal = {AAPPS bulletin},
	publisher = {Springer},
	author = {Lau, Jonathan Wei Zhong and Lim, Kian Hwee and Shrotriya, Harshank and Kwek, Leong Chuan},
	year = {2022},
	pages = {27},
}

@article{peruzzo_variational_2014,
	title = {A variational eigenvalue solver on a photonic quantum processor},
	volume = {5},
	number = {1},
	journal = {Nature communications},
	publisher = {Nature Publishing Group UK London},
	author = {Peruzzo, Alberto and McClean, Jarrod and Shadbolt, Peter and Yung, Man-Hong and Zhou, Xiao-Qi and Love, Peter J and Aspuru-Guzik, Alán and O’brien, Jeremy L},
	year = {2014},
	pages = {4213},
}

@article{hakim_quantum_2026,
	title = {Quantum {Slater}-exponent optimization for electronic-structure calculations on simulated noisy intermediate-scale quantum computers},
	volume = {113},
	number = {2},
	journal = {Physical Review A},
	publisher = {APS},
	author = {Hakim, Difa Farhani and Prayitno, Teguh Budi and Choong, Pak Shen and Sarwono, Yanoar Pribadi and Kwek, Leong-Chuan},
	year = {2026},
	pages = {022610},
}

@article{arute_quantum_2019,
	title = {Quantum supremacy using a programmable superconducting processor},
	volume = {574},
	number = {7779},
	journal = {nature},
	publisher = {Nature Publishing Group UK London},
	author = {Arute, Frank and Arya, Kunal and Babbush, Ryan and Bacon, Dave and Bardin, Joseph C and Barends, Rami and Biswas, Rupak and Boixo, Sergio and Brandao, Fernando GSL and Buell, David A and {others}},
	year = {2019},
	pages = {505--510},
}

@article{darmawan_improving_2025,
	title = {Improving qubit reduction for molecular simulations with randomized orbital sampling},
	volume = {35},
	number = {1},
	journal = {AAPPS Bulletin},
	publisher = {Springer},
	author = {Darmawan, Yoga A and Fauzi, Angga D and Sarwono, Yanoar P and Zhang, Rui-Qin},
	year = {2025},
	pages = {27},
}

@article{iryanti_connecting_2025,
	title = {Connecting ansatz design to density functional observables using weighted subspace search vqe for kohn-sham hamiltonian},
	volume = {100},
	number = {12},
	journal = {Physica Scripta},
	publisher = {IOP Publishing},
	author = {Iryanti, Ira and Hue, Jun Hao and Soesanto, Qidir Maulana Binu and Hanna, M Yusrul and Rangkuti, Choirun Nisaa and Chuan, Kwek Leong and Sarwono, Yanoar P},
	year = {2025},
	pages = {125414},
}

@article{bharti_noisy_2022-1,
	title = {Noisy intermediate-scale quantum algorithm for semidefinite programming},
	volume = {105},
	number = {5},
	journal = {Physical Review A},
	publisher = {APS},
	author = {Bharti, Kishor and Haug, Tobias and Vedral, Vlatko and Kwek, Leong-Chuan},
	year = {2022},
	pages = {052445},
}

@article{kandala_hardware-efficient_2017,
	title = {Hardware-efficient variational quantum eigensolver for small molecules and quantum magnets},
	volume = {549},
	number = {7671},
	journal = {nature},
	publisher = {Nature Publishing Group UK London},
	author = {Kandala, Abhinav and Mezzacapo, Antonio and Temme, Kristan and Takita, Maika and Brink, Markus and Chow, Jerry M and Gambetta, Jay M},
	year = {2017},
	pages = {242--246},
}

@article{seeley_bravyi-kitaev_2012,
	title = {The {Bravyi}-{Kitaev} transformation for quantum computation of electronic structure},
	volume = {137},
	number = {22},
	journal = {The Journal of chemical physics},
	publisher = {AIP Publishing},
	author = {Seeley, Jacob T and Richard, Martin J and Love, Peter J},
	year = {2012},
}

@article{patel_quantum_2025,
	title = {Quantum measurement for quantum chemistry on a quantum computer},
	volume = {125},
	number = {16},
	journal = {Chemical Reviews},
	publisher = {ACS Publications},
	author = {Patel, Smik and Jayakumar, Praveen and Yen, Tzu-Ching and Izmaylov, Artur F},
	year = {2025},
	pages = {7490--7524},
}

@article{ikhtiarudin_measurement_2026,
	title = {Measurement cost reduction in {ADAPT}-{VQE} via variance-based shot allocation and {Pauli}-measurement reuse},
	volume = {101},
	number = {25},
	journal = {Physica Scripta},
	publisher = {IOP Publishing},
	author = {Ikhtiarudin, Azhar and Sunnardianto, Gagus Ketut and Fathurrahman, Fadjar and Agusta, Mohammad Kemal and Dipojono, Hermawan Kresno},
	year = {2026},
	pages = {255103},
}

@article{zhu_optimizing_2024,
	title = {Optimizing shot assignment in variational quantum eigensolver measurement},
	volume = {20},
	number = {6},
	journal = {Journal of Chemical Theory and Computation},
	publisher = {ACS Publications},
	author = {Zhu, Linghua and Liang, Senwei and Yang, Chao and Li, Xiaosong},
	year = {2024},
	pages = {2390--2403},
}

@article{mariyanto_fubini-study_2026,
	title = {Fubini-{Study} {Metric} {Tensor} {Evolution} and {Reduced} {State} {Distinguishability} in {Geometry}-{Aware} {Optimization} for {Reliable} {Near}-{Term} {Variational} {Quantum} {Eigensolvers}},
	volume = {126},
	number = {13},
	journal = {International Journal of Quantum Chemistry},
	publisher = {Wiley Online Library},
	author = {Mariyanto, Dwi Cahyo and Prihadi, Hadyan L and Fauzi, Angga Dito and Handoko, LT and Sarwono, Yanoar P and Zhang, Rui-Qin},
	year = {2026},
	pages = {e70255},
}

@article{romero_strategies_2019,
	title = {Strategies for quantum computing molecular energies using the unitary coupled cluster ansatz},
	volume = {4},
	number = {1},
	journal = {Quantum Science and Technology},
	publisher = {IOP Publishing},
	author = {Romero, Jonathan and Babbush, Ryan and McClean, Jarrod R and Hempel, Cornelius and Love, Peter J and Aspuru-Guzik, Alán},
	year = {2019},
	pages = {014008},
}

@article{verteletskyi_measurement_2020,
	title = {Measurement optimization in the variational quantum eigensolver using a minimum clique cover},
	volume = {152},
	number = {12},
	journal = {The Journal of chemical physics},
	publisher = {AIP Publishing},
	author = {Verteletskyi, Vladyslav and Yen, Tzu-Ching and Izmaylov, Artur F},
	year = {2020},
}

@book{casella_statistical_2024,
	title = {Statistical inference},
	publisher = {CRC press},
	author = {Casella, George and Berger, Roger},
	year = {2024},
}

@article{barkoutsos_quantum_2018,
	title = {Quantum algorithms for electronic structure calculations: {Particle}-hole {Hamiltonian} and optimized wave-function expansions},
	volume = {98},
	number = {2},
	journal = {Physical Review A},
	publisher = {APS},
	author = {Barkoutsos, Panagiotis Kl and Gonthier, Jerome F and Sokolov, Igor and Moll, Nikolaj and Salis, Gian and Fuhrer, Andreas and Ganzhorn, Marc and Egger, Daniel J and Troyer, Matthias and Mezzacapo, Antonio and {others}},
	year = {2018},
	pages = {022322},
}

@article{tilly_variational_2022,
	title = {The variational quantum eigensolver: a review of methods and best practices},
	volume = {986},
	journal = {Physics Reports},
	publisher = {Elsevier},
	author = {Tilly, Jules and Chen, Hongxiang and Cao, Shuxiang and Picozzi, Dario and Setia, Kanav and Li, Ying and Grant, Edward and Wossnig, Leonard and Rungger, Ivan and Booth, George H and {others}},
	year = {2022},
	pages = {1--128},
}

@incollection{kais_introduction_2014,
	title = {Introduction to {Quantum} {Information} and {Computation} for {Chemistry}},
	isbn = {978-1-118-74263-1},
	url = {https://onlinelibrary.wiley.com/doi/abs/10.1002/9781118742631.ch01},
	doi = {https://doi.org/10.1002/9781118742631.ch01},
	booktitle = {Quantum {Information} and {Computation} for {Chemistry}},
	publisher = {John Wiley \& Sons, Ltd},
	author = {Kais, Sabre},
	year = {2014},
	note = {Section: 1},
	pages = {1--38},
}

@article{lanyon_towards_2010,
	title = {Towards quantum chemistry on a quantum computer},
	volume = {2},
	number = {2},
	journal = {Nature chemistry},
	publisher = {Nature Publishing Group},
	author = {Lanyon, Benjamin P and Whitfield, James D and Gillett, Geoff G and Goggin, Michael E and Almeida, Marcelo P and Kassal, Ivan and Biamonte, Jacob D and Mohseni, Masoud and Powell, Ben J and Barbieri, Marco and {others}},
	year = {2010},
	pages = {106--111},
}

@article{kassal_simulating_2011,
	title = {Simulating chemistry using quantum computers},
	volume = {62},
	number = {1},
	journal = {Annual review of physical chemistry},
	publisher = {Annual Reviews},
	author = {Kassal, Ivan and Whitfield, James D and Perdomo-Ortiz, Alejandro and Yung, Man-Hong and Aspuru-Guzik, Alán},
	year = {2011},
	pages = {185--207},
}

\end{document}